\documentclass{article}

\usepackage{arxiv}

\usepackage[utf8]{inputenc} 
\usepackage[T1]{fontenc}    
\usepackage{hyperref}       
\usepackage{url}            
\usepackage{booktabs}       
\usepackage{amsfonts}       
\usepackage{nicefrac}       
\usepackage{microtype}      
\usepackage{lipsum}
\usepackage{graphicx}
\usepackage{fixltx2e}
\graphicspath{ {./images/} }

\title{Classification Of Sleep-Wake State In A Ballistocardiogram System Based On Deep Learning}

\author{
 Nemath Ahmed \\
  Turtle Shell Technologies Pvt. Ltd.\\
  Bangalore, India  \\
  \texttt{nemath.ahmed@dozee.io} \\
   \And
 Aashit Singh \\

  Turtle Shell Technologies Pvt. Ltd.\\
  Bangalore, India  \\
  \texttt{aashit@dozee.io} \\
  \And
 Srivyshnav KS \\

  Turtle Shell Technologies Pvt. Ltd.\\
  Bangalore, India  \\
  \texttt{srivyshnav@dozee.io} \\
    \And
 Gulshan Kumar \\

  Department of Neurophysiology\\
  National Institute of Mental Health\\ and Neurosciences(NIMHANS)\\
  Bangalore, India  \\
  \texttt{gulshankumar@nimhans.edu.in} \\
    \And
 Gaurav Parchani\\
 
  Turtle Shell Technologies Pvt. Ltd.\\
  Bangalore, India  \\
  \texttt{gaurav@dozee.io} \\
    \And
 Vibhor Saran \\

  Turtle Shell Technologies Pvt. Ltd.\\
  Bangalore, India  \\
  \texttt{vibhor@dozee.io} \\
}

\begin{document}
\maketitle
\begin{abstract}
 Sleep state classification is vital in managing and understanding sleep patterns and is generally the first step in identifying acute or chronic sleep disorders. However, it is essential to do this without affecting the natural environment or conditions of the subject during their sleep.  Techniques such as \textbf{P}oly\textbf{s}omno\textbf{g}raphy\textbf{(PSG)} are obtrusive and are not convenient for regular/long-term sleep monitoring. Fortunately, The rise of novel technologies and advanced computing has given a recent resurgence to many sleep-monitoring techniques. One such contactless and unobtrusive monitoring technique is \textbf{B}allisto\textbf{c}ardio\textbf{g}raphy \textbf{(BCG)}, in which vitals are monitored by measuring the body's reaction to the cardiac ejection of blood. In this study, we propose a Multi-Head 1D-Convolution based Deep Neural Network to classify sleep-wake state and predict sleep-wake time accurately using the signals coming from a BCG sensor. Our method achieves a sleep-wake classification score of 95.5\%, which is on par with recent works based on the PSG system. We further conducted two independent studies in a controlled and uncontrolled environment to test the sleep time and wake up time prediction accuracy. We achieve a score of 94.16\% in a controlled environment on 115 subjects and 94.90\% in an uncontrolled environment on 350 subjects. The high accuracy and contactless nature of the proposed system make it a convenient method for long term monitoring of sleep states.

\end{abstract}


\section{Introduction}
Sleep-Awake state detection refers to identifying and classifying sleep and awake episodes for a subject. Sleep-state estimation is one of the first steps in analyzing and approaching solutions to more serious sleep disorders.  Most sleep disorders remain largely undiagnosed in the general population \cite{Roth2002-oh} due to lack of continuous monitoring techniques. Many of these sleep disorders can be identified using sleep state estimation and analyzing sleep patterns \cite{Cho2019-uh} . Techniques to build novel solutions for efficient estimation of sleep-state have been one of the important topics of focus for the past few decades \cite{Kulkarni2019-qn, Palotti2019-tk}. Today researchers are leveraging the high computing power and modern AI-based techniques \cite{Krizhevsky2017-wn,Dean_Jeffrey_and_Corrado_Greg_and_Monga_Rajat_and_Chen_Kai_and_Devin_Matthieu_and_Mao_Mark_and_Ranzato_Marcaurelio_and_Senior_Andrew_and_Tucker_Paul_and_Yang_Ke_and_others2012-da,Kingma2014-cz,Esmaeilzadeh2020-fd,Nadarzynski2019-hm,Han2020-dl,Kaur2018-hv,Jiang2017-on} to achieve better results. Earlier researches focused on traditional techniques such as Polysomnography (PSG) which is regarded as the current gold standard in high-resolution sleep monitoring, but the method is expensive and recordings are performed in an unfamiliar and controlled environment \cite{Estrada2004-tt, Jorgensen2020-ee}. To achieve PSG data, individuals typically have to spend the night in a sleep laboratory—a controlled setting under the continued supervision of a sleep technician. With the rise of computational efficiencies, researchers started exploring wearables for detecting sleep-state \cite{Yuzer2020-sa,Boe2019-xs,Dhamchatsoontree2019-si,Jean-Louis2001-sk}. Techniques such as \textbf{W}rist \textbf{A}ctigraphy \textbf{(WA)} \cite{Martin2011-ts,Jean-Louis2001-sk,Liang2019-nc} where devices worn around the wrist record movements that can estimate sleep parameters with specialized algorithms were introduced to compute total sleep time, efficiency, and intermediate awakenings . These devices are easy to use as compared to burdensome PSG  which employs electrodes to measure brain dynamics of EEG, eye movements, muscle activity, heart physiology, and respiratory function \cite{Martin2011-ts,Marino2013-ld}. However, as compared with PSG, actigraphy is known to overestimate sleep and underestimate the awake time and is known to lack critical confirmation including standardization for device settings. The device can be uncomfortable when worn all the time making continuously monitoring difficult. In this study, we propose an efficient contactless sleep-state monitoring using ballistocardiography (BCG) and deep learning. 

Non-contact ballistocardiography (BCG) is an unobtrusive and non-invasive system that evaluates cardiovascular functions without any difficulty. In comparison to PSG, ballistocardiography does not require external physical electrodes to be connected and avoids any direct contact to a subject, avoiding any uneasiness and discomfort. Such a system is suitable for discreet long-term continuous data acquisition \cite{Giovangrandi2011-qg,Saran2018-rw}. 

With the rise of computing power, there have been tremendous advances in the field of \textbf{A}rtificial \textbf{I}ntelligence (\textbf{AI}) \cite{Nadarzynski2019-hm,Han2020-dl,Kaur2018-hv,Jiang2017-on}.  These tools may improve prognosis, diagnostics, and care planning and it is believed that AI will be an integral part of healthcare services in the near future \cite{Esmaeilzadeh2020-fd}. In this study, we focus on deep learning \cite{ Esmaeilzadeh2020-fd,Krizhevsky2017-wn,Dean_Jeffrey_and_Corrado_Greg_and_Monga_Rajat_and_Chen_Kai_and_Devin_Matthieu_and_Mao_Mark_and_Ranzato_Marcaurelio_and_Senior_Andrew_and_Tucker_Paul_and_Yang_Ke_and_others2012-da,Kingma2014-cz} based on \textbf{C}onvolutional \textbf{N}eural \textbf{N}etworks (\textbf{CNNs}) which have come to be recognized as prominent feature extractors. For medicine in general, CNNs have enabled Computer-Aided Diagnosis to occasionally outperform experts \cite{Han2020-dl,Hwang2019-ya,McKinney2020-xy}.  In this study, we explore CNN based techniques and analyze their performance on non-contact BCG data. We use multi-head 1D-CNN architecture to classify the sleep state and a prediction algorithm is run to obtain sleep and wake-up time. We also discuss the real-time implementation of this approach along with the integration of transfer learning for long term stability.

\section{Background}
\label{sec:headings}
Ballistocardiogram (BCG) captures the reaction of the whole body resulting from the cardiac ejection of blood. In other words, it is a combined effect of the movement of blood inside the heart, arteries, and movement of the heart itself. It also captures the movements generated by the body during respiration and other bodily movements. With the recent rise of novel technologies and advanced signal processing methods, ballistocardiography has noticed a recent resurgence \cite{Giovangrandi2011-qg}. Originally invented in the late 19th century \cite{Gordon1877-gz}, BCG became a topic of interest around the 1940s and continued to the late 20th century. The reason for its disappearance thereafter is attributed to lack of standard measurement techniques, high noise, lack of understanding of the physiologic origin, lack of computing to perform complex operations and the rise of ECG and ultrasound which were far more accurate than BCG. In the last decade, there has been a massive surge \cite{Giovangrandi2011-qg,Saran2018-rw,Scarborough1956-tb,Inan2009-pl,Cathelain2020-xg,Mora2020-uo,Saran2019-yq} in focus around BCG to address and propose solutions to the shortcomings of this archaic method. One of the major advantages of using BCG as a vital monitoring technique is that no trained professionals are required to carry out procedures like the placement of electrodes, unlike ECG. BCG can become an alternative to ECG due to its easy-to-use non-contact method of monitoring a subject’s vitals.

\subsection{Works in BCG}
With the motive to reduce the complexities in monitoring human-vitals, many researchers have come up with novel techniques to put BCG into best use. In \cite{Saran2018-rw} a novel algorithm to detect individual heartbeats from BCG data with a higher detection rate and accuracy was proposed. The proposed algorithm was able to achieve an accuracy of 98.39\%\ to capture heart rate with a detection rate of 99.46\%\ compared to a standard ECG machine. G. Cathelain \textit{et al.} \cite{Cathelain2020-xg} proposed a neural network to learn inherent physiological variability and artifacts in the BCG data previously unattainable through signal processing techniques. The proposed U-Net architecture in the study was able to detect heartbeats with 92\%\ precision. Neural Networks have also previously as well been used to detect and classify heartbeats as in \cite{Camps2018-bl}. The advent of AI has thoroughly helped in the introduction and development of such techniques.

\subsection{Other Recent Works}
 With regard to the importance of sleep-state detection for further detection of sleep disorders, popular studies have come up with introducing novel techniques to accurately detect sleep-state. Most of these studies are based on PSG, EEG \cite{Kulkarni2019-qn,Saleab2016-fj}, ECG, or Actigraphy \cite{Boe2019-xs,Dhamchatsoontree2019-si,Jean-Louis2001-sk} based techniques that involve a certain level of discomfort when considered for daily monitoring. A deep learning approach \cite{Kulkarni2019-qn} to efficiently detect sleep-spindles was proposed based on a single EEG channel. It was able to do better than the then state-of-the-art algorithm because of good generalization obtained by using deep neural networks. This also forms an inspiration to our work around the use of deep learning as the sleep patterns are diverse and vary drastically from person to person. In  M.S. Saleab \textit{et al.} \cite{Saleab2016-fj} proposed a real-time sleep detection and warning system based on EEG. The proposed system achieved an accuracy of 96.3\%\, 100\%\ of sensitivity, 92.4\%\ of Predictability, and 93\%\ Specificity. In G. Jean-Louism \textit{et al.} \cite{Jean-Louis2001-sk} we see the comparison between accelerometer-based actigraph with PSG. However, both the study involved a controlled environment and hence concluding with the scores that are difficult to reproduce in normal/uncontrolled conditions.

 In \cite{Yuzer2020-sa,Boe2019-xs} we see the introduction of wearables in place of traditional EEG and PSG taking shape. A.J. Boe \textit{et al.} \cite{Boe2019-xs} introduces wireless, wearable sensors that were used in automatic sleep classification. The proposed system was however tested on a small subject cluster and resulted in 74.4\%\ specificity and 90.0\%\ sensitivity which are far away from medical grade scores. This method of tracking vitals through wearables might prove useful for people with skin allergies or burn patients. In one of the more recent works by S. Dhamchatsoontree \textit{et al.} \cite{Dhamchatsoontree2019-si}, we see the use of pressure sensors backed by machine learning algorithms to give better Sleep Quality Index (SQI) and sleep postures. However,  it was able to reach accuracies above 87\%\ in detecting SQI and sleep postures.

 In many works\cite{Cho2019-uh,Jean-Louis2001-sk,Liang2019-nc} we also see a combination of wearables backed by machine learning and deep learning techniques such as LSTMs, CNNs, and RNNs. T.Cho \textit{et al.} \cite{Cho2019-uh} proposed a 3-axis accelerometer-based wristband and an end-to-end deep learning model (Deep-ACTINet) to produce results accurate of up to 89.65\%\ in sleep-state classification. The algorithm used CNNs and LSTM to produce outputs that represent sleep-state. In  work by G. Jean-Louis \textit{et al.} \cite{Jean-Louis2001-sk}, again we see the use of LSTM but here the dataset was much stronger, comprising 186 subjects. An interesting approach was proposed where in addition to physiological data, behavioral data from mobile phones was also used. The algorithm achieves higher performance in real-life ambulatory settings. S. Biswal \textit{et al.} \cite{Biswal2017-ah} proposed an automated sleep staging method via deep learning with the introduction of SleepNet architecture. The accuracy of the study on 10,000 patients was 85.76\%\ on average. The work by J. Palottu \textit{et. al} \cite{Palotti2019-tk} goes on to show the potential of CNNs with regards to actigraphy techniques.

In this work we try to show how a contactless technique like Ballistocardiography can be on par with Polysomnography when backed by strong deep learning architecture. In this paper, we propose :
\begin{itemize}
\item The use of contact-less Ballistocardiography technique for efficient detection of sleep-state.
\item Proposing a new Multi-Head CNN based architecture for classification.
\item A method for predicting sleep time and wake-up time based on Multi-Head classification.
\end{itemize}

\begin{figure} 
    \centering
    \includegraphics{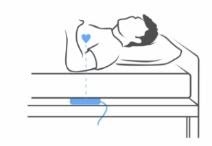}
     \caption{Dozee BCG sensor in use}
     \label{dozee_in_use}
\end{figure}

\section{Methodology}
\subsection{Data}
\subsubsection{BCG Data Acquisition}
We use Dozee \cite{Saran2018-rw,Saran2019-yq}, a contactless sleep and body vitals monitoring device (shown in Figure \ref{dozee_in_use}) for acquiring BCG data. The system comprises of a mesh of Polyvinylidene fluoride (PVDF) - based vibroacoustic sensors placed under the mattress to capture micro and macro-vibrations generated by the body which includes cardiac contractions, breathing, body movements, snoring when lying over the sensor array is attached to a data-acquisition unit sampling vibrations at a rate of 250Hz. All the recordings were done in accordance with the Declaration of Helsinki and informed consent form was obtained from all the subjects in the study.

\subsubsection{Features}
 In this study, we use five parameters to build the feature matrix that is then feeded to the Multi-Head CNN architecture for classification. These five features are 1) Heart Rate, 2) Breath Rate, 3) Heart Rate Confidence (confidence of calculated heart rate), 4) Movements(micro/macro movements captured by Dozee), and 5) Difference in successive Heart Rates. These parameters have shown a great influence in the prediction of sleep time and wake up time of a subject. All these features are obtained using the Dozee system \cite{Saran2018-rw,Saran2019-yq}.

\begin{figure} 
    \centering
    \resizebox{\textwidth}{!}{\includegraphics{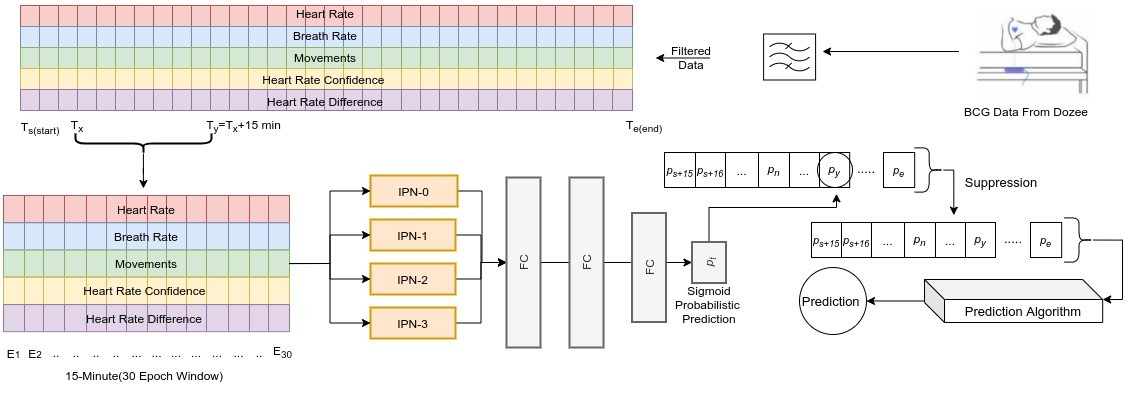}}
     \caption{The complete flow of the algorithm is shown in the figure above. The filtered data from Dozee(BCG Sensor Sheet) is divided into a 15-minute length rolling window and sent through a classifier. The final classification scores for the complete data are sent to a prediction algorithm. The prediction algorithm returns predicted sleep time and wake-up time.}
     \label{pipeline}
\end{figure}

\subsection{Sleep-State Classification}
\subsubsection{CNNs}
In the last few years, Convolutional Neural Networks (CNNs), a deep learning architecture based on the neurobiological structure in the visual cortex of mammals has achieved state-of-the-art performances in many domains. Introduced by LeCun \textit{et al} \cite{LeCun1995-lx} in 1995, it’s use has also been seen in multiple types of research around sleep-state and sleep-disorders \cite{Cho2019-uh,Jean-Louis2001-sk,Liang2019-nc}. CNNs comprise a class of neural network architecture that uses a series of convolutional filters, non-linear activation functions, and pooling layers to minimize a loss function. We use CNNs as an efficient feature extractor for extracting complex features and relations in the BCG data acquired from Dozee BCG sensors.

In this study, we focus on 1D CNNs on our time series data. As shown in Fig \ref{pipeline}, the BCG data is sampled such that we get a value for each epoch, each epoch being equivalent to 30 seconds. We use 15-minute (30 epochs) windows as input for classifying the 15th minute as sleep or awake. We perform the prediction on a rolling window of length 15-minutes (30 epochs) separated by 1-minute (2 epochs). We run our prediction algorithm on classification scores to find the probable sleep time and probable wake-up time over the data available from BCG sensors. We perform this for the entire duration where data from BCG sensors is available and give predictions in real-time.

\subsubsection{1D CNNs}
 Studies in \cite{Kiranyaz2015-px,Kiranyaz2016-sj,Kiranyaz2017-cr,Kiranyaz2019-up,Kiranyaz2019-ze,Wu2020-fm,Avci2017-dz,Avci2018-dw,Abdeljaber2017-cz,Avci2018-yr,Ince2016-mt,Abdeljaber2018-uh} have shown how 1D CNN might be advantageous in comparison to traditional 2D CNNs which generally are used with images and videos. 1) 
1D CNNs are less computationally complex in comparison to 2D CNNs due to simple array operations instead of matrix operations. 2) Studies have shown how 1D CNNs have shallow architectures allowing them to learn challenging tasks involving 1D signals. 2D CNNs on the other hand require deeper architectures to learn similar types of features and are more expensive memory-wise. 3) Due to their low computational requirements, compact 1D CNNs are well-suited for real-time and low-cost applications especially on mobile or hand-held devices \cite{Kiranyaz2015-px}. These features make 1D CNNs as the perfect choice in this study.
\begin{figure} 
    \centering
    \includegraphics[scale=0.5]{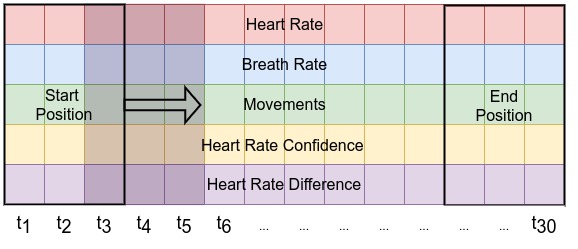}
     \caption{Input feature matrix for the Multi-Head CNN architecture.}
     \label{1DCNN}
\end{figure}
As mentioned earlier, we have five features of focus which are Heart Rate, Breath Rate, Movements, Heart Confidence, and Difference in successive heartbeat intervals(Heart Rate Difference).  In the 1D CNN approach, a feature detector is slid from the start position till the end position extracting deep features at each instance (Fig \ref{1DCNN}).

The width of the feature detector shows the resolution at which the detector looks at the whole 15-minutes (30 epochs) moving from left to right as in Fig \ref{1DCNN}.  In Fig \ref{1DCNN}, we have a detector width of the length of 3 epochs or 6 minutes meaning the first feature extraction takes place from all the feature values between t\textsubscript{1}-t\textsubscript{3}, the next from t\textsubscript{2}-t\textsubscript{4}, and so on till t\textsubscript{28}-t\textsubscript{30}.  

\subsubsection{Multi-Head Network}
The vanilla 1-D CNN neural network is rigid and captures features at a defined resolution. Since there is a strong relation between signal amplitudes at various time instances with the sleep-state, it becomes necessary to look at the 15-minute window with different resolutions. The idea was to make a prediction based on combined information obtained from sliding the window at different resolutions. 

\begin{figure} 
    \centering
    \includegraphics[scale=0.5]{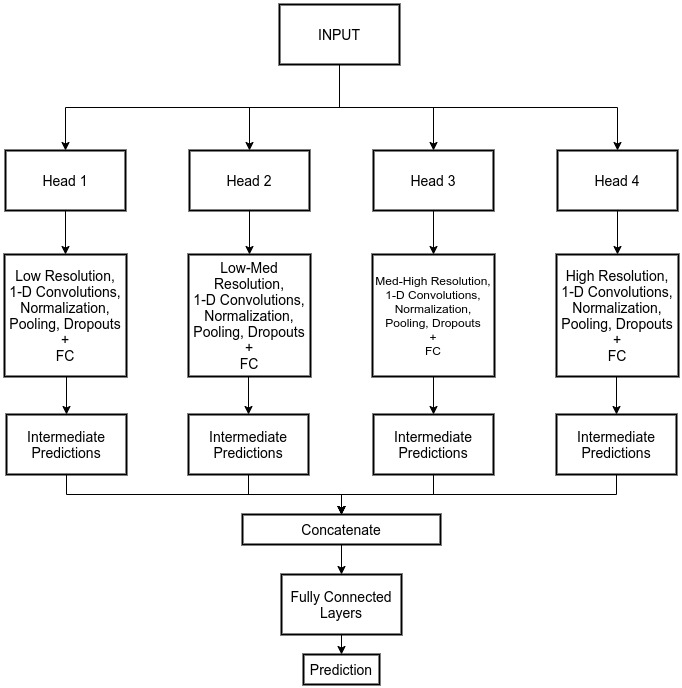}
     \caption{CNN based Multi-Head Network.}
     \label{multihead}
\end{figure}

We thus use four different heads and process each of them  at resolutions of three, five, seven and eleven epochs respectively. The features obtained through these 1D convolutions are then passed on to an intermediate \textbf{F}ully-\textbf{C}onnected (\textbf{FC}) layer in order to make intermediate predictions. Before passing on to the FC layers, we perform batch normalization, low-level max-pooling, and apply dropouts in order to avoid overfitting.  The four intermediate predictions obtained after processing by intermediate FC layers correspond to predictions based on four different resolutions. These intermediate predictions are then concatenated and passed on to a simple shallow fully-connected network whose output is probabilistic binary prediction. The out-line architecture is given in Figure \ref{multihead}.
\subsubsection{Prediction Algorithm}
On a rolling basis, we get the output of Multi-Head CNN as the probability of sleep and awake state. We feed this into our prediction algorithm which forms the final step in obtaining sleep-time and wake-up time. We first pass the prediction probability array through a 1-D Non-Max Suppression algorithm which eliminates any small period sleep-state changes helping in better prediction. After this, the probabilities are converted to binary symbols and passed through the prediction algorithm. The prediction algorithm in simple terms finds the probable sleep time and probable wake-up time and gives them an output based on a few conditions. 

We consider a probable sleep time timestamp as sleep time if there are at least 45-minutes of sleep observed after the probable-sleep-time timestamp with no 10 minutes of continuous awake predictions. Similarly, a time stamp is considered as predicted wake-up time if there is at least a 15-minute of awake prediction observed before this probable-wakeup timestamp and none observed later. We arrived at these conditions based on numerous tests performed on above 1000 periods of sleep from 350 subjects. This is the final step and the output of this process gives the final sleep-time and wake-up time prediction.

\section{Results}
\subsection{Evaluation Method}
 For training, we use BCG data collected from 250 subjects in an uncontrolled setting amounting to an average of 6.5 hours of data per user. This data was then annotated using feedback from participants and markings by experts and was divided into multiple 15-minute windows of sleep and awake. Using the data from 250 subjects, we were able to produce 100,000 15-minute windows of sleep and awake which where used as input to the deep learning architecture. From the complete data available for a subject, we considered the data one hour before and one hour after the sleep-state change.  The training was done on NVidia RTX 2060 with 6-Gigabyte of memory and was also trained to CPU for avoiding discrepancies while deploying the model to a CPU-based server. All the results are the average scores that were obtained after ten independent runs.

\begin{table}
\parbox{.45\linewidth}{
\centering
\caption{Classification Techniques}
\begin{tabular}{cc}
\hline
    Model        & Classification Score \\
    \midrule
    \textbf{Model-1(MH 1D CNNs)} & \textbf{95.5\%}     \\
    Model-2(1D CNNs) & 90.8\%\     \\
    Model-3(2D CNNs)  &86.7\%\   \\
    Model-4(MT 1D  CNNs) & 84.3\%\ \\
    \bottomrule
\hline
\end{tabular}
\label{class_tech}

}
\hfill
\parbox{.45\linewidth}{
\centering
\caption{State Classification (Controlled)}
\begin{tabular}{cc}
\hline
    Measure        &  Score \\
    \midrule
    Accuracy & 92.63\%\     \\
    Precision & 95.80\%\     \\
    Specificity  &95.49\%\   \\
    Sensitivity & 90.12\%\ \\
    \bottomrule
\hline
\end{tabular}
\label{state_classification_controlled}
}
\end{table}
As discussed, the first step of our method is classifying the 15-minute window as a sleep window or an awake window  on a roll of one-minute for the entire duration of data. Before arriving at Multi-Head 1D convolutions, multiple other approaches were experimented with a goal to achieve maximum accuracy in classifying the 15-minute window. Classification accuracy has a significant effect on the final outcome (sleep-wake time) and therefore this first step is of prime importance. The approaches which were experimented are mentioned in Table \ref{class_tech}. The scores are based on the classification of 20,000 15-minute windows as sleep or awake. 
 In Model-2 a similar architecture to a single \textbf{I}ntermediate \textbf{P}rediction \textbf{N}etwork (\textbf{IPN}) was used. Though the model was of low computational costs, the accuracy was not up to the mark.  Model-3 was based on converting the BCG data to interpretable signal images and using 2-D convolutions to extract the features based on signals introduced in previous sections. The model proved to be less efficient and was of high computational cost. In Model-4 a multi-task network based on 1D CNNs was built and it proved to be both inaccurate and computationally expensive. Model-1 proved to be the most efficient and accurate though being computationally a little expensive compared to Model-2.

   We follow two procedures for testing our method. First, we perform tests on subjects in a controlled environment using verification from PSG and then we extend this to testing our method in an uncontrolled setting. Both of these evaluation procedures prove that our method is consistent with PSG methodology and that it is well suited even in a normal (uncontrolled) environment such as at home-based monitoring.

\begin{table}
\parbox{.45\linewidth}{
\centering
\caption{Sleep-Wake Time Prediction(controlled)}
\begin{tabular}{cc}
\hline
    Measure        &  Score \\
    \midrule
    Accuracy & 94.16\%\     \\
    Precision & 97.32\%\     \\
    Specificity  &97.14\%\   \\
    Sensitivity & 91.53\%\ \\
    \bottomrule
\hline
\end{tabular}
\label{S/W-time-cont}

}
\hfill
\parbox{.45\linewidth}{
\centering
\caption{Sleep-Wake Time Prediction(uncontrolled)}
\begin{tabular}{cc}
\hline
    Measure        &  Score \\
    \midrule
    Accuracy & 94.90\%\     \\
    Precision & 96.04\%\     \\
    Specificity  &96.13\%\   \\
    Sensitivity & 93.67\%\ \\
    \bottomrule
\hline
\end{tabular}
\label{S/W-time-uncon}
}
\end{table}
\subsubsection{Evaluation 1- Controlled  Environment} 
   In this evaluation, we score our method based on the medical data gathered from PSG of 115 subjects. All the recordings were done using Nihon Kohden Neurofax EEG-1200 machine \cite{noauthor_1958-xf} (24-bit resolution, 1024 Hz sampling rate and 0.1-250Hz bandpass filter) with 24 electrodes (19-EEG:Electroencephalography electrodes as per Jasper’s 10-20 \cite{Berry2012-rv} system, 2-EOG: Electro-oculography and 3-EMG:Electromyography electrodes as per the guidelines of American Academy of Sleep Medicine  \cite{noauthor_2012-jk}. This evaluation proves the alignment and consistency of our method to the standard PSG techniques. We match each of our 15-minute classifications as sleep or awake with the PSG data. We use the evaluation parameters mentioned below for checking performance. The scores obtained are tabulated in Table \ref{state_classification_controlled}. 
\[ Accuracy =\frac{TP+TN}{P+N}\times100 \ \ 
\ \ Precision=\frac{TP}{TP+FP}\times100 \] 
\[ Specificity =\frac{TN}{FP+TN}\times100 \ \ 
\ \ Sensitivity=\frac{TP}{TP+FN}\times100 \] 
   A \textbf{T}rue \textbf{P}ositive(\textbf{TP}) is an outcome where the model correctly predicts the positive class. Similarly, a \textbf{T}rue \textbf{N}egative(\textbf{TN}) is an outcome where the model correctly predicts the negative class.A \textbf{F}alse \textbf{P}ositive(\textbf{FP}) is an outcome where the model incorrectly predicts the positive class and \textbf{F}alse \textbf{N}egative(\textbf{FN}) is an outcome where the model incorrectly predicts the negative class.

  The results show that our method gives an accuracy of 92.6\% in the classification of a 15-min window which is even higher than most of the actigraphy based sleep awake prediction techniques\cite{Jean-Louis2001-sk,Liang2019-nc}. This proves the efficacy of the use of Multi-Task 1D CNN based network for sleep-wake classification and sleep time and wake-up time prediction.  We then score our method based on sleep and wake up timestamps. We take into account a 15-min relaxation time between the actual time and the predicted time. The scores are summarised in the Table \ref{S/W-time-cont}.

\subsubsection{Evaluation-2 Uncontrolled Environment} 
    In this evaluation, we perform tests on 350 subjects with ground truths being based on the feedback from users. These 350  subjects were tested on multiple random days resulting in a total of 1500 samples averaging to approx. 7 hrs of data per sample. The final scores are shown in Table \ref{S/W-time-uncon}. The method gives high sensitivity, specificity, and precision scores which are higher than most wearable actigraphy techniques\cite{Jean-Louis2001-sk,Yuzer2020-sa,Boe2019-xs,Dhamchatsoontree2019-si,Liang2019-nc}. The method also comes close to PSG techniques which are currently regarded as the gold standard.This goes on to evaluate the BCG as an effective way of easy non-contact sleep monitoring of a person even from the comfort of their home.
    
\subsection{Transfer Learning}
   The proposed method has a small margin of error which can be eliminated by the use of transfer learning. Transfer learning is the improvement of learning in a new task through the transfer of knowledge from a related task that has already been learned \cite{Jialin2009-hp,Bengio2012-qj,Taylor2009-tz}. As the vitals vary from person to person it is difficult to generalize a common deep learning model for all the subjects. We thus propose an extended transfer learning module that can be developed for people belonging to different buckets of sleep patterns. For example, a bucket of people with a periodic limb movement disorder, a bucket of people with constant sleep fidgeting, and so on. 

The learning is done on the fully connected layers after the concatenation of intermediate prediction networks. The remaining weights in the architecture are frozen as these are precious features learned from over 100,000 samples. We test this approach on 10 subjects out of the 350 subjects mentioned in Evaluation-1 for whom the prediction failed due to distinct pattern vitals when compared to the general trend. 

   Samples from 5 random days were collected for each of the users to perform transfer learning on the original architecture. For testing, samples from 10 days distinct from the training samples were collected and accuracy was measured both by our base architecture and by the transfer learning approach. There was an average increase of 22\% inaccuracy using the new model based on transfer learning compared to the base model. Also, a maximum increment of 40\% was observed in the new model. This shows that when done at a broader scale, this can further reduce the error in our approach.

\section{Discussion}
 One of the most important steps before analyzing sleep-related disorders is to accurately determine the clinical awakeness of the subject. Even in long term healthcare monitoring, sleep patterns can give us good insights into the health of a person and their mental well-being. This makes the topic of sleep awake classification a topic of interest to many researchers. 
   We can see a lot of research \cite{Kulkarni2019-qn,Saleab2016-fj,Yuzer2020-sa,Boe2019-xs,Dhamchatsoontree2019-si,Jean-Louis2001-sk,Kim2018-fh,Liang2019-nc,Biswal2017-ah,Palotti2019-tk} which shows that this topic has been a subject for a number of prior research.  However, most of them alter the natural sleeping conditions by involving monitoring methods that are intrusive, uncomfortable, and impractical for daily usage. M.S. Saleab \textit{et. al} \cite{Saleab2016-fj} proposed a method which gives an accuracy of 96.3\%\ accuracy in prediction of sleep/awake stage. However, the study involves the use of many sensors such as pulse, oxygen saturation, body temperature, galvanic response, and few other sensors. In \cite{Yuzer2020-sa} we see the use of accelerometers placed on a person’s diaphragm. Boe \textit{et. al} \cite{Boe2019-xs} introduced a multimodal sensor system measuring hand acceleration, electrocardiography, and distal skin temperature detecting awake and sleep with a recall of 74.4\%\ and 90.0\%\, respectively which are considerably less compared to our method. The method is also tedious and cannot be used for daily monitoring. The method proposed has been evaluated on 115 subjects in controlled and 350 subjects in an uncontrolled settings. High accuracy and efficiency demonstrated makes this method a prime contender for usage in hyper-scaled uncontrolled settings.
Most of the studies around sleep awake detection have focused on a small study group in a medically controlled environment which brings out the robustness of our approach. The contactless technique with good accuracy, evaluation, and generalization makes our method novel. This method can be further extended to:
\begin{itemize}

\item Assessment of insomnia and excessive daytime somnolence
\item Assessment of sleep apnea and respiratory disturbances
\item Assessment of transient and chronic schedule disorders 
\item Sleep in psychiatric and other medical conditions
\item Assessment of medical interventions-drug studies
\end{itemize}

\section{Conclusion}
In this study, we propose a Deep Learning based Multi-Head architecture for the classification of sleep-awake state and we extend it to sleep time wake up time prediction with our proposed prediction algorithm. We use 1-D Convoultional Neural Networks in our Multi-Head architecture to extract features and patterns in the data coming from BCG sensor. We show how the results from the proposed method with non-contact BCG are comparable to other contact based techniques. We do this by evaluating our state classification and time prediction on subjects both in controlled and uncontrolled environments. The results obtained are also at-par with other contact-based actigraphy techniques which makes our proposed non-contact approach desirable. This study can be further extended to sleep stage classification in future research.

\section{Acknowledgement}
All the validation PSG studies were conducted at the Human Sleep Research Laboratory of the Department of Neurophysiology at NIMHANS. We would like to thank Dr. Bindu M Kutty and her team at the lab for the help they provided in conducting these studies and acquiring validation data. This work is partly funded under the Biotechnology Innovation Grant Scheme of the Government of India.

\bibliographystyle{unsrt}  
\bibliography{references}

\end{document}